# Apparel Recommender System based on Bilateral image shape features

Yichi Lu[1], Mingtian Gao[1], Ryosuke Saga[1]

[1]Osaka Prefecture University, Japan
(Tel: 81-72-254-9678, Fax: 81-72-254-9678)

[1]saga@cs.osakafu-u.ac.jp

**Abstract:** Probabilistic matrix factorization (PMF) is a well-known model of recommender systems. With the development of image recognition technology, some PMF recommender systems that combine images have emerged. Some of these systems use the image shape features of the recommended products to achieve better results compared to those of the traditional PMF. However, in the existing methods, no PMF recommender system can combine the image features of products previously purchased by customers and of recommended products. Thus, this study proposes a novel probabilistic model that integrates double convolutional neural networks (CNNs) into PMF. For apparel goods, two trained CNNs from the image shape features of users and items are combined, and the latent variables of users and items are optimized based on the vectorized features of CNNs and ratings. Extensive experiments show that our model predicts outcome more accurately than do other recommender models.

## 1 INTRODUCTION

The sparseness of user–item ratings in e-commerce services is a major data type. Traditional recommendation systems have focused on predicting the rating accuracy on the basis of several supplemental features.

The target of the current study is apparel. Apparel is different from other goods, such as books and movies. Visual information is more important than document information because, generally, the latter is useful for improving recommendation accuracy [5]–[11]. The most well-known approaches are probabilistic matrix factorization (PMF) [1] and singular value decomposition. However, these models are often faced with sparse matrices, such that they fail to extract effective representations from users and items. Several models have been proposed to overcome the above problem. Wang et al. proposed collaborative deep learning that integrates PMF and deep learning to learn hidden representation [11]. However, this method utilizes bag-of-work, which cannot treat contextual information. Kim et al. addressed this problem and proposed the combination of convolutional matrix factorization (ConvMF) with PMF and convolutional neural network (CNN) [6].

With the development of deep learning and its widespread use in various fields, the use of image shape features has a positive effect on recommender systems. This study draws on other bilateral models to modify ISFMF into a bilateral model and obtain better results. It proposes a framework called bilateral image shape feature probabilistic matrix factorization (Bi-ISFMF), which has two image shape features for CNN to collaborate with PMF. Extensive experiments show that our model predicts the outcome more accurately than do other recommender models.

## 2 PRELIMINARIES

### 2.1 PMF

Salakhutdinov et al. [1] proposed the PMF, which is a well-known approach for recommendation systems. We suppose that $M$ users, $N$ items, and a $R \in R^{N \times M}$ rating matrix exist. We also require the user latent matrix $U \in R^{k \times N}$ and item latent matrix $V \in R^{k \times M}$ to reconstruct the rating matrix $R$. The goal of PMF is to determine the optimal matrix $U$. $V$ minimizes the loss function $\mathcal{E}$, as shown in the following:

$$min\mathcal{E}(U,V) = \sum_i^N \sum_j^M \frac{I_{ij}}{2}(r_{ij} - u_i^T v_j)^2 \\ + \frac{\lambda_U}{2}\sum_i^N ||u_i||^2 + \frac{\lambda_V}{2}\sum_j^M ||v_j||^2 \quad (1)$$

### 2.2 Deep Features

In the last few years, deep learning has made significant progress in natural language processing, image processing, and object detection, and high-level features have been extracted with the use of deep neural networks. These features are also being applied to recommender systems [6].

Kim et al. [6] addressed the limitations of the bag-of-words model-based approaches and proposed a novel document context-aware recommendation model (i.e., ConvMF) that integrates CNN into PMF. The proposed model's CNN can capture the contextual meaning of words in documents and even distinguish subtle contextual differences of the same word via different shared weights. Duan et al. [8] used the deep feature of the product image to improve the recommendation accuracy of the apparel recommender system. This deep feature is extracted by DeepContour. This method was proposed by Shen et al. [15], and compared to the traditional method, it can extract more accurate image shape features. The authors used CNN in learning contour features to improve the accuracy of contour detection. They divided the contour data into subclasses based on the contour shape, thereby converting the contour versus non-contour classification problem into a multi-class classification problem.

## 3 BI-ISFMF

### 3.1 Probabilistic Model of Bi-ISFMF

Bi-ISFMF is the modified model from ISFMF. In Bi-ISFMF, the image shape features of products previously purchased by customers and of recommended products are used simultaneously. Fig. 1 shows the overview of the probabilistic model of Bi-ISFMF. First, suppose that $N$ users, $M$ items, and a user-item rating matrix ($R \in R^{N \times M}$) exist. In a probabilistic point of view, the conditional probability of the observed rating matrix $R$ is expressed as follows:

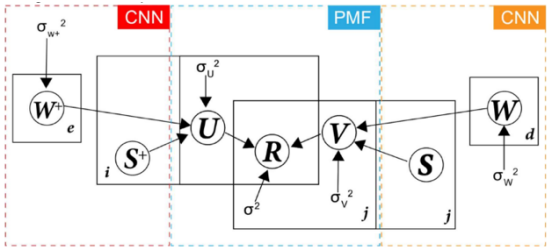

**Fig. 1.** Graphical model of Bi-ISFMF

$$p(R \mid U, V, \sigma^2) = \prod_i^N \prod_j^M \left[ N(r_{ij} \mid u_i^T v_j, \sigma^2) \right]^{I_{ij}} \quad (2)$$

where $N(X \mid \mu, \sigma^2)$ is a Gaussian distribution of $X$ with mean $\mu$ and variance $\sigma^2$. As for the users' latent factors, we assume that a latent factor is formed by three variables: $W^+$ which is internal weights in CNN, $S_i^+$ which is an image of item $i$, and $\varepsilon_i$ which is Gaussian noise of item $i$, i.e. $\varepsilon_i \sim N(0, \sigma_i^2 I)$. By using these variables, the users' latent factors $u_i$ is expressed as

$$u_i = cnn(W^+, S_i^+) + \varepsilon_i \quad (3)$$

where $cnn()$ represents the output of our CNNs' architecture. A zero-mean spherical Gaussian prior, the most commonly used prior, is placed for each weight $W_e^+$ in $W^+$.

$$p(W^+ \mid \sigma_{w^+}^2) = \prod_e Nw(_e^+ \mid 0, \sigma_{w^+}^2) \quad (4)$$

Accordingly, the conditional distribution over users' latent factors is given by

$$p(U \mid W^+, S^+, \sigma_U^2) = \prod_i^N N(u_i \mid cnn(W^+, S_i^+), \sigma_U^2 I) \quad (5)$$

Just like the user's latent factor, the item's latent factor consists of three variables, $W$: internal weights $W$ in right CNN, $S_j$ which is an image of item $j$, and $\varepsilon_j$ which is variable of the Gaussian noise, i.e. $\varepsilon_j \sim N(0, \sigma_j^2 I)$. And by using three variables, the item's latent factor shows

$$v_j = cnn(W, S_j) + \varepsilon_j \quad (6)$$

where $cnn()$ represents the output of our CNNs' architecture. A zero-mean spherical Gaussian prior, the most commonly used prior, is placed for each weight $w_d$ in $W$.

$$p(w \mid \sigma_w^2) = \prod_d N(w_d \mid 0, \sigma_w^2) \quad (7)$$

Accordingly, the conditional distribution over items' latent factors is given by

$$p(V \mid W, S, \sigma_V^2) = \prod_j^M N(v_j \mid cnn(W, S), \sigma_V^2 I) \quad (8)$$

### 3.2. Architecture of our CNNs

The image of item $i$ (multiple images) bought by the user in the past is processed by DeepContour [15] to extract the image shape features. We use transfer learning to place a pre-trained CNN model (DeepContour) into our structure for contour detection. This pre-trained model has four convolutional layers and three fully connected layers. The output of the first fully connected layer is considered for our proposed model. Next, a 45 × 45 three-channel image patch is used as input and a 128-dimensional feature vector of the image shape is obtained. These latent vectors are then concatenated and compressed into a vector of dimension k, which is passed to the PMF and incorporated into matrix decomposition. The second CNN is for extracting image features from recommended products. Different from the previous CNN, it extracts only one image per recommended item from each user. Thus, this CNN has no concatenate layer.

In addition, theoretically, the mechanism is designed to compress all images into a vector of dimension k if the graphics card has sufficient memory to support it.

### 3.3. Optimization

To optimize the variables, we use maximum a posteriori estimation as follows:

$$\max_{U,V,w^+,w} p(U, V, W^+, W \mid R, S^+, S, \sigma^2, \sigma_U^2, \sigma_V^2, \sigma_{W^+}^2, \sigma_W^2)$$
$$= \max_{U,V,w^+,w} [p(R \mid U, V, \sigma^2) p(U \mid W^+, S^+, \sigma_U^2) p(W^+ \mid \sigma_{W^+}^2) \quad (9)$$
$$p(V \mid W, S, \sigma_V^2) p(W \mid \sigma_W^2)]$$

If a negative logarithm is given on Equation (10), it can be reformulated as follows:

$$min\mathcal{E}(U, V, W^+, W) = \sum_i^N \sum_j^M \frac{I_{ij}}{2} (r_{ij} - u_i^T v_j)^2$$
$$+ \frac{\lambda_U}{2} \sum_i^N ||u_i| - cnn(W^+, S_i^+)|^2 + \frac{\lambda_{W^+}}{2} \sum_e^{|W_e^+|} ||W_e^+||^2 \quad (10)$$
$$+ \frac{\lambda_V}{2} \sum_j^M ||v_j| - cnn(W, S_j)|^2 + \frac{\lambda_W}{2} \sum_d^{|W_d|} ||W_d||^2$$

In which $\lambda_U$ is $\sigma^2/\sigma_U^2$, $\lambda_V$ is $\sigma^2/\sigma_V^2$, $\lambda_{W^+}$ is $\sigma^2/\sigma_{W^+}^2$, and $\lambda_W$ is $\sigma^2/\sigma_W^2$.

A coordinate descent is adopted that iteratively optimizes a latent variable while fixing the remaining variables. Specifically, Equation (11) becomes a quadratic function with respect to $U$ (or $V$) while temporarily assuming $W$ and $V$ (or $W^+$ and $U$) can be analytically computed in a closed form by simply differentiating the optimization function $min\mathcal{E}$ with respect to $U_i$ (or $V_j$) as follows:

$$u_i = (VI_iV^T + \lambda_U I_K)^{-1}(VR_i + \lambda_U cnn(W^+, S_i^+)) \quad (11)$$

$$v_j = cnn(W, S_j) + \varepsilon_j \tag{12}$$

where $I_i$ is a diagonal matrix with $I_{ij}$ $j = 1, \dots, M$ is its diagonal elements, and $R_i$ is a vector with $(r_{ij})_{i-1}^{M}$ for user $i$. For item $j$, $I_j$ and $R_j$ are similar defined as $I_i$ and $R_i$, respectively. Unfortunately, $W^+$ and $W$ are closely related to the feature of CNN architecture, such as max-pooling layers and nonlinear activation functions.

Fortunately, when $U$ and $V$ are temporarily fixed, loss function $\varepsilon$ becomes an error function with regularized terms of neural net.

$$\begin{aligned}\varepsilon(W^+) &= \frac{\lambda_U}{2}\sum_i^N ||u_i| - cnn(W^+, S_i^+)||^2 \\ &+ \frac{\lambda_{W^+}}{2}\sum_e^{|w_e^+|} ||W_e^+||^2 + \text{constant}\end{aligned} \tag{13}$$

$$\begin{aligned}\varepsilon(W) &= \frac{\lambda_V}{2}\sum_j^M \| v_j| - cnn(W, S_j)|^2 \\ &+ \frac{\lambda_W}{2}\sum_d ||W_d||^2 + \text{constant}\end{aligned} \tag{14}$$

To optimize $W^+$ and $W$, we use the backpropagation algorithm with a given target value $V_j$, which is temporary fixed. The overall optimization process ($U$, $V$, $W^+$ and $W$ are alternatively update) is repeated until convergence. With the optimized $U$, $V$, $W^+$ and $W$, finally, the rating of user $i$ on item $j$ can be predicted as follows:

$$\begin{aligned}\hat{r}_{ij} &\approx \mathbb{E}[r_{ij} \mid u_i^T v_j, \sigma^2] = u_i^T v_j \\ &= (cnn(W^+, S_i^+) + \varepsilon_i)^T (cnn(W, S_j) + \varepsilon_j)\end{aligned} \tag{15}$$

## 4 EXPERIMENT

### 4.1. Goal, Dataset, Environments, and Criteria

In this experiment, our proposed model is evaluated for apparel. Clothes and accessories from the Amazon product dataset are used as the category data. The dataset consists of rating data and the images and documents of items. We pre-processed the dataset for the experiment as follows:

• The dataset is divided into two, namely, clothes and accessories, to investigate the performance of our model for different product types.

• Items that do not have their images are removed from the two datasets.

• The matrix size of the training data is set similarly to that of the original data to divide the training and test data. Users who have less than two ratings are removed. The statistics of each datum validate that the two datasets have different characteristics (Table 1.).

We compare the Bi-ISFMF with the following baselines:

• PMF [1]: PMF is a standard rating prediction model for user ratings only.

• ConvMF [6]: This model was introduced by Kim et al. It uses the contextual features of items to improve the accuracy of rating prediction.

• ISFMF [18]: This model uses transfer learning to integrate DeepContour into the model proposed by Duan et al.

**Table 1.** Detail statistics of the two datasets

| Dataset | Users | Items | Ratings | Density |
|---|---|---|---|---|
| Clothes | 8222 | 8513 | 19654 | 0.0281% |
| Accessories | 3889 | 3507 | 8544 | 0.0626% |

### 4.2. Experiment Setup

Our experimental environment uses the Keras Python library with NVIDIA GeForce Titan Xp and NVIDIA GeForce 1080. The parameter settings are as follows:

• The size of the latent dimension of U and V is set at 50.
• The size of the input image is set at 60 × 60.
• The number of input images is between 2 and 5.
• Each dataset is randomly split into training, validation, and test sets.

The hyperparameters ($\lambda_U, \lambda_V$) are selected via grid search for different models, and these hyperparameters may vary. Table II lists the best combination of hyperparameters ($\lambda_U, \lambda_V$). Root mean square error (RMSE) is used for the evaluation measure as follows:

$$\text{RMSE} = \sqrt{\frac{\sum_{i,j}^{N,M}(r_{ij} - \hat{r}_{ij})^2}{ratings}} \tag{16}$$

With regard to the size of the input image, we tested different image sizes and calculated that, though no significant difference was found, the size is best set at 60 × 60.

### 4.3. Experimental Results

Table 2. shows the overall RMSE of the PMF, ConvMF, ISFMF, Left-ISFMF, and Bi-ISFMF with different numbers of input images on the dataset. Imp denotes the improvement, wherein our model outperforms the ISFMF. In comparison with the other four models, Bi-ISFMF attains significant performance on the dataset. The improvement of our model over the best competitor, namely, ISFMF, increases consistently from 5.12% to 7.23% and from 9.76% to 11.99% on the Clothes dataset and the Accessories dataset, respectively. Left-ISFMF has a similar performance as ConvMF. However, when the training set is 80%, only the accuracy of the Bi-ISFMF decreases. No specific correlation appears between the accuracy and the number of input images. In general, the larger the quantity, the higher the accuracy in this model. An additional experiment was conducted to verify this observation.

Table 2. Overall RMSE test on Dataset

| Model | Clothes | | Accessories | |
|---|---|---|---|---|
| | 70 | 80 | 70 | 80 |
| PMF | 1.646 | 1.603 | 1.729 | 1.633 |
| ConvMF | 1.327 | 1.272 | 1.466 | 1.433 |
| ISFMF | 1.162 | 1.152 | 1.334 | 1.322 |
| Bi-ISFMF | 1.080 | 1.093 | 1.174 | 1.197 |
| imp | 7.23% | 5.12% | 11.99% | 9.76% |

The calculation is repeated 100 times to determine the average for the clothes dataset with different numbers of images. Accuracy increases as the number increases, but not much growth is observed. The average value is also lower but the variance is larger.

## 5 CONCLUSION

This study determined whether the use of bilateral image shape features can effectively improve the rating prediction accuracy of an apparel recommendation system. Owing to the limitations of the experimental environment, we could not calculate experiments with more than five input images for each user. The experiment results indicated that, though accuracy will improve with the increase in the number of input images, it will stabilize eventually. Hence, it is not necessary to input more images for Bi-ISFMF.

For future works, different visual information will be used to investigate the accuracy of the recommendation systems.